\documentclass[aps,pra,10pt,twocolumn,numerical]{revtex4-1}
\usepackage[utf8x]{inputenc}
\usepackage{amssymb,amsmath}
\usepackage{multirow}
\usepackage{graphicx}
\usepackage{epsfig}
\usepackage{hyperref}
\usepackage{subfigure}
\usepackage{bbm}
\usepackage{stmaryrd}
\usepackage{pspicture}
\usepackage{natbib}
\usepackage[english]{babel}
\usepackage{overpic}
\usepackage{epstopdf}

\usepackage{hyperref}
\hypersetup{colorlinks=true, breaklinks=true, linkcolor=blue, citecolor=red}

\newcommand{\op}[1]{\hat{#1}^{\phantom{\dagger}}}
\newcommand{\opd}[1]{\hat{#1}^{\dagger}}
\newcommand{\tr}[1]{{\rm{Tr}}\left\{#1\right\}}
\newcommand{\ie}{i.e.}
\newcommand{\eg}{e.g.}

\newcommand{\ii}{{\rm{i}}\,}

\newcommand{\equ}[1]{\begin{align}{#1}\end{align}}

\newcommand{\rklamm}[1]{ \left( #1 \right) }
\newcommand{\eklamm}[1]{ \left[ #1 \right] }
\newcommand{\sklamm}[1]{ \left\{ #1 \right\} }

\newcommand{\figref}[1]{Fig.~\ref{#1}}
\newcommand{\secref}[1]{Sec.~\ref{#1}}
\newcommand{\equref}[1]{Eq.~\eqref{#1}}

\begin{document}
\title{Nonequilibrium relaxation transport of ultracold atoms}
\author{Fernando Gallego-Marcos}
  \email{fergallmar@hotmail.com}
  \affiliation{Instituto de Ciencia de Materiales, CSIC, Cantoblanco, 28049 Madrid, Spain}
\author{Christian Nietner} 
  \email{cnietner@itp.tu-berlin.de}
  \affiliation{Institut f\"ur Theoretische Physik, Technische Universit\"at Berlin, Hardenbergstr. 36, 10623 Berlin, Germany}
\author{Gernot Schaller}
  \affiliation{Institut f\"ur Theoretische Physik, Technische Universit\"at Berlin, Hardenbergstr. 36, 10623 Berlin, Germany}
\author{Tobias Brandes}
  \affiliation{Institut f\"ur Theoretische Physik, Technische Universit\"at Berlin, Hardenbergstr. 36, 10623 Berlin, Germany}
\author{Gloria Platero}
  \affiliation{Instituto de Ciencia de Materiales, CSIC, Cantoblanco, 28049 Madrid, Spain}
  
%
%
%
\begin{abstract}
  We analyze the equilibration process between two either fermionic or bosonic reservoirs containing ultracold atoms with a fixed total number of particles that are weakly connected via a few-level quantum system. We allow for both the temperatures and particle densities of the reservoirs to evolve in time. Subsequently, linearizing the resulting equations enables us to characterize the equilibration process and its time scales in terms of equilibrium reservoir properties and linear-response transport coefficients. Additionally, we investigate the use of such a device as particle transistor or particle capacitor and analyze its efficiency.\\[.3cm]
  PACS number(s): 67.85.-d, 51.10.+y, 05.60.Gg, 05.30.Rt
\end{abstract}
\maketitle
%
%
\section{Introduction}
%
%
Transport phenomena are of utmost importance in a whole variety of scientific research fields such as biology, chemistry, and physics. Here, systems which are initialized in nonequilibrium strive to equilibrate with their surrounding by exchanging energy and particles until a stationary state is reached. This equilibration is quite well understood for classical systems, where it usually results in a thermal steady state. However, despite its importance, relaxation and thermalization in closed quantum systems are still not fully understood \cite{Goldstein2006,Linden2009,Yukalov2011,Reimann2012,Eisert2012}.

In recent years it has become possible to isolate and control quantum systems to a very high degree. Namely, there has been a lot of progress in the production and manipulation of ultracold quantum gases in ultrahigh vacuum chambers, using optomagnetical traps \cite{Zoller1998,Courteille2001} and lasers \cite{Greiner2002,Jordens2008,Ott2008}. Here, the system of interest is isolated from its environment to such a high degree that thermodynamic variables are not tunable externally, but are solely determined implicitly by the system itself. Therefore, in such systems it is necessary to calculate the thermodynamic variables self-consistently in order to correctly describe their equilibration properties.

Thereby, the quantum mechanical peculiarities become relevant and potentially observable and measurable in an experiment. This has been impressively confirmed by the creation of the famous Bose-Einstein condensate \cite{Anderson1995,Davis1995}. After successfully studying setups with ultracold atoms in equilibrium configurations for quite a while, nowadays, the focus shifts to investigating their nonequilibrium properties \cite{Mandel2003,Palzer2009,Hurtado2011,Ates2012,Chien2012a,Chien2012b,Wacker2013,Chien2013,Salger2013,Ventra2013}. However, measuring the properties of such systems is quite complicated and usually results in the destruction of the system. A possible evasion of this problem could be the observation of transport processes, as has been also theoretically researched for setups involving atomic reservoirs coupled to, \eg, each other \cite{Ponomarev2011}, a lattice system \cite{Bruderer2012}, a potential trap \cite{Gutman2012}, or even quantum dot systems \cite{Recati2005,Ivanov2013,Nietner2013}
. 
Following this idea, recent experiments \cite{Krinner2013,Brantut2012,Brantut2013}, which investigate the transport properties between two ultracold atomic clouds, are especially noteworthy.

Motivated by these experiments, we analyze within this paper a transport setup consisting of a mesoscopic few-level quantum system in contact with two ultracold particle reservoirs, whose thermodynamic variables are calculated self-consistently. We explicitly include a few-level system in our model, since it enhances the quantum character of the transport setup, as is well known from electronic and photonic mesoscopic transport, where one observes effects such as the Kondo effect \cite{Hewson,Meir1992,Chen2004,Chang2005}, Coulomb blockade \cite{Kouwenhoven1997,Kouwenhoven1997b,Grabert1992,Nguyen2007}, coherent population trapping \cite{Groth2006,Busl2010}, and dark states \cite{Brandes2000,Michaelis2006,Emary2007}, to name but a few. Furthermore, this approach, in principle, allows for an external control of the equilibration process via the few-level quantum system.

In \secref{S:ThoereticalFramework}, we start by presenting the general theoretical framework which we use throughout this paper. Here, we first review the properties of ideal quantum gases in \secref{S:AtomicReservoirs}, and in \secref{S:MasterEquation}, we derive the master-equation formalism which we use to describe the transport through an open quantum system. In \secref{S:EoM}, we deduce the resulting system of equations of motion and additionally establish a linear-response theory in \secref{S:LinearResponse}. Subsequently, we apply this formalism to different setups and present the respective results in \secref{S:Results}. In particular, we investigate fermionic systems with one and two transition energies in Secs.~\ref{sec::fersingle} and \secref{sec::fertwo}, respectively. In comparison, we additionally analyze a bosonic system with two transition energies in \secref{sec::bosetwo}. Finally, we summarize our results in \secref{S:Summary}.

Note that throughout this paper we use the natural units with $\hbar=k_B=1$.
%
%
\section{Theoretical Framework}\label{S:ThoereticalFramework}
%
%
In real experiments with cold atoms the chemical potential can not be tuned directly by applying an external voltage as usually considered for electronic transport. Instead, one can introduce a thermal or density gradient which causes a bias in the chemical potentials of the reservoirs.

In order to describe this bias correctly, we need to determine the chemical potential self-consistently from the reservoir temperature and particle number. Therefore, we review the properties of ideal quantum gases within the next subsection.
%
%
\subsection{Ideal Quantum Gases}\label{S:AtomicReservoirs}
%
%
We model the ultracold atomic baths as ideal, noninteracting quantum gases of spinless massive particles trapped in a three-dimensional (3D) inhomogeneous harmonic potential \cite{Jana2012} with the effective trapping volume 
\equ{
    \bar\omega^3\equiv\omega_x\omega_y\omega_z, 
    }
resulting from the trapping frequencies along the $x$, $y$, and $z$ directions.

The reservoirs are described by the total Hamiltonian $\hat{\mathcal H}_{\rm{B}}=\sum_\nu\hat{\mathcal H}^{(\nu)}_{\rm{B}}$, where the Hamiltonian for each connected bath $\nu \in \{L,R\}$ is given by
\begin{equation}\label{E:BathHamiltonian}
	\hat{\mathcal H}^{(\nu)}_{\rm{B}}=\underset{\boldsymbol n}{\sum}\varepsilon_{\boldsymbol n}\,\hat{b}_{\nu,\boldsymbol n}^{\dagger}\,\hat{b}_{\nu,\boldsymbol n},
\end{equation}
with either bosonic or fermionic operators $\hat{b}_{\nu,\boldsymbol n}^{\dagger}$ and $\hat{b}_{\nu,\boldsymbol n}$, which create and annihilate a particle in the quantum state $\boldsymbol n = (n_x,n_y,n_z)^{\rm T}$ with energy $\varepsilon_{\boldsymbol n} = (\boldsymbol n+1/2) \boldsymbol \omega$ in reservoir $\nu$ with trapping frequencies $\boldsymbol \omega = (\omega_x,\omega_y,\omega_z)^{\rm T}$. These baths are weakly coupled to the system via the interaction Hamiltonian
\begin{equation}\label{E:SysBathHamiltonian}
	\hat{\mathcal H}_{\rm{SB}}=\underset{\nu,\boldsymbol n}{\sum} \left( t_{\nu,\boldsymbol n} \, \hat{b}_{\nu,\boldsymbol n}^{\dagger}\,\hat{a} + \rm{H.\,c.} \right),
\end{equation}
where the operators $\op{a}$ and $\opd{a}$ annihilate and create particles in the few-level quantum system. Here, the tunneling amplitude of an atom hopping from the reservoir $\nu$ into the system or vice versa is proportional to $t_{\nu,\boldsymbol n}^{*}$ and $t_{\nu,\boldsymbol n}$, respectively.

In what follows , we parametrize the tunneling amplitudes by energy-dependent tunneling rates formally defined by
\equ{
  \Gamma_\nu(\omega)=\sum_{\boldsymbol n} 2\pi \left|t_{\nu,\boldsymbol n}\right|^2 \delta(\omega-\varepsilon_{\boldsymbol n}).\label{E:TunnelingRates}
    }
Assuming that the reservoirs equilibrate sufficiently fast, at each point in time, they can approximately be characterized by their equilibrium distributions $\bar n_\nu^{(\xi)} (\varepsilon_{\boldsymbol n}) = 1/\eklamm{e^{\beta_\nu(\varepsilon_{\boldsymbol n} -\mu_\nu)} - \xi}$, where $\xi=+1$ corresponds to a Bose gas and $\xi=-1$ corresponds to a Fermi gas. Here, we  introduce the inverse temperature $\beta_\nu=1/T_\nu$ and the chemical potential $\mu_\nu$ of each reservoir.

Furthermore, we can derive the macroscopic equilibrium variables $T_\nu$, $\mu_\nu$, and $N_\nu$ of the reservoirs in the grand canonical ensemble, using the condition that the average number of particles $N_\nu =\sum_{\boldsymbol n} \bar n_\nu^{(\xi)}(\varepsilon_{\boldsymbol n}) $ is constant.
With this, we obtain the well-known expressions for the average number of particles $N_\nu$ confined in a harmonic trapping potential \cite{Pethick,Bagnato2005}
\begin{align}
     { N_\nu} &= \xi \rklamm{\frac{T_\nu}{\bar\omega}}^{3}  {\rm{Li}}_{{3}}(\xi z_\nu)+ N_\nu^{(0)}{(\xi)},\label{E:AverageParticleNumber}
\end{align}
where the correction to the number of particles in the ground state is given by
\begin{align}
     N_\nu^{(0)}{(\xi)} &= \begin{cases}
	\frac{z_\nu}{1-z_\nu}&:\xi=+1,\\
	0&:\xi=-1,
	\end{cases} \label{E:GoundStateParticleNumber}
\end{align}
and the average internal energy reads as
\equ{
    {U_\nu} &= 3\, \xi \,\bar \omega  \rklamm{\frac{T_\nu}{\bar\omega}}^{4} {\rm{Li}}_{4}(\xi z_\nu)\label{E:InternalEnergyGeneral}.
}

Here, we introduced the fugacity $z_\nu=\exp({\beta_\nu \mu_\nu})$ and the polylogarithm ${\rm{Li}}_s (x)= \sum_{k=1}^\infty x^k /k^s$ \cite{Abramowitz}.

Note that the additional ground-state contribution $N_\nu^{(0)}(\xi)$ is only present for bosonic gases. Since \equref{E:AverageParticleNumber} implicitly defines the chemical potential $\mu_\nu= \mu_\nu(T_\nu,N_\nu)$ as a function of temperature and mean particle number, the ground-state contribution from \equref{E:GoundStateParticleNumber} can be associated to the phenomenon of Bose-Einstein condensation, where the chemical potential vanishes and the occupation of the ground state becomes macroscopic \cite{Pitaevskii}. This critical behavior is characterized by a corresponding critical temperature $T_{C} =\bar\omega \eklamm{N/\zeta(3) }^{1/3}$, with the Riemann zeta function $\zeta(s)$. In the fermionic case, the Fermi temperature $T_{F} =\bar \omega\eklamm{ \frac{4}{3}N/\zeta(3)}^{1/3}$ characterizes the ideal Fermi gas.

Also, notice that the harmonic trapping potential is solely chosen due to its experimental relevance. The general method proposed within this paper is equivalently applicable to other confinement potentials such as, \eg, a 3D cubic box. Then, Eqs.~\eqref{E:AverageParticleNumber} and \eqref{E:InternalEnergyGeneral} need to be changed accordingly (see Appendix \ref{A:Box}).
\begin{figure}[t]
 \centering
 \includegraphics[width=.95\columnwidth]{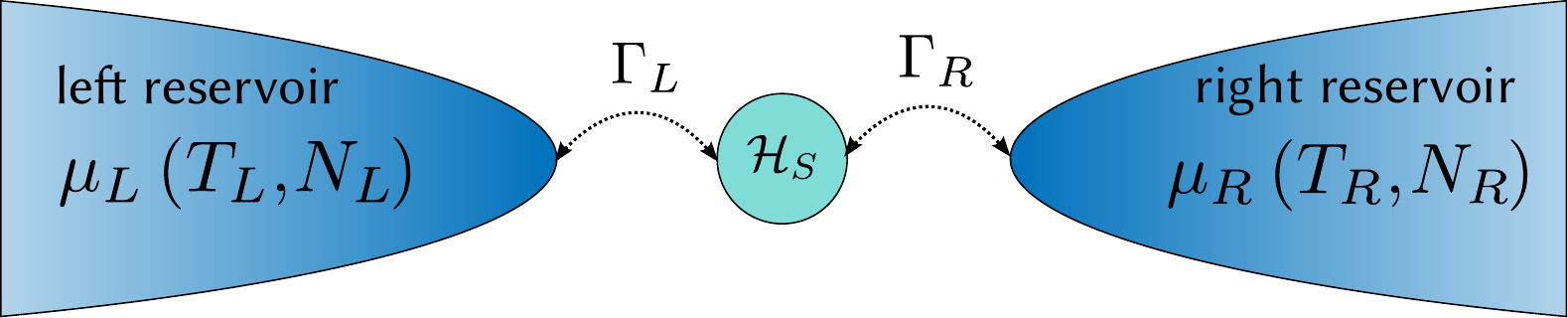}
 \caption{(Color online) General two-terminal transport scheme with left and right reservoir weakly coupled to a few-level quantum system. The reservoirs $\nu \in \sklamm{ L , R }$ are in thermal equilibrium and characterized by a chemical potential $\mu_\nu(T_\nu,N_\nu)$ that depends on the respective temperature $T_\nu$ and particle number $N_\nu$. The system dynamics is governed by the Hamiltonian $\mathcal{H}_S$ and the weak system-bath coupling is facilitated by energy-dependent tunneling rates $\Gamma_\nu(\omega)$ defined in \equref{E:TunnelingRates}. }\label{F:GeneralTransportScheme}
\end{figure}
%
%
\subsection{Transport Master Equation}\label{S:MasterEquation}
%
%
We investigate a general transport setup as sketched in \figref{F:GeneralTransportScheme}, with two reservoirs, denoted by the labels $L$ and $R$, which are independently in thermal equilibrium and coupled to the transport system. We assume that the system-bath coupling is sufficiently weak, \ie, $\Gamma_\nu(\omega)\ll k_B T_\nu$, such that we can make use of the Born-Markov secular approximation (BMS) \cite{Breuer}.

Starting from the von Neumann equation
  $\dot\varrho=-\ii [\hat{\mathcal H},\varrho]$
of the full system $\hat{\mathcal H} = \hat{\mathcal H}_{\rm{S}} + \hat{\mathcal H}_{\rm{B}}+ \hat{\mathcal H}_{\rm{SB}}$ with the full density matrix $\varrho$, this formalism allows one to extract a quantum master equation, which assumes the form of a rate equation for the reduced system density matrix $\rho={\rm{Tr_{B}}}\sklamm{\varrho}$ in the system energy eigenbasis for nondegenerate energy eigenvalues \cite{SchallerBook}. Here, ${\rm{Tr_B}}\{ \textrm\textbullet \}$ represents a trace over all bath degrees of freedom.

Consequently we obtain a rate equation for the populations of the reduced system density matrix which reads as
\equ{
    \dot \rho_i(t) =\sum_\nu \sum_{j} \mathcal L_{ij}^{(\nu)}(t) \rho_j(t)\label{eq::rhofermi},
    }
where $\rho_i$ represents the population of the $i$th system energy eigenstate and the summation runs over all energy eigenstates $j$ of the system Hamiltonian $\hat{\mathcal{H}}_{\rm S}$ and all attached reservoirs $\nu$. The rate matrix elements satisfy the condition $\sum_{i}\mathcal L_{ij}^{(\nu)} (t)=0$.
Identifying the jump terms in the master equation \eqref{eq::rhofermi}, we can deduce the energy and matter currents which for sequential tunneling are given by
\equ{
    \dot N_\nu (t)&=\sum_{j>i}\eklamm{\mathcal L_{ij}^{(\nu)}\rho_j-\mathcal L_{ji}^{(\nu)}\rho_i},\label{E:StaticCurrentDefinition}\\
    \dot E_{\nu} (t)&=\sum_{j>i} (\omega_j-\omega_i)\eklamm{\mathcal L_{ij}^{(\nu)}\rho_j-\mathcal L_{ji}^{(\nu)}\rho_i}\label{E:StaticEnergyCurrentDefinition},
    }
where $\omega_i$ is the eigenenergy of the few-level quantum system corresponding to the population $\rho_i$. Here, we defined the currents in such a way that they are negative if the particles flow from the reservoir $\nu$ into the few-level quantum system.

These currents are parametrized by the energy-dependent tunneling rates $\Gamma_\nu(\omega)$ defined in \equref{E:TunnelingRates} and, in general, depend on the equilibrium quantum statistics $\bar n_\nu^{(\xi)} (\omega)$ of the attached reservoirs. 

Moreover, it is convenient to further combine the energy and particle currents and introduce the heat current flowing from reservoir $\nu$ into the system as \cite{Esposito2009b,Esposito2010b}
\equ{
      \dot {Q}_\nu(t) \equiv \dot E_\nu (t)- \mu_\nu (t)\dot N_\nu(t). \label{E:HeatCurrent}
    }

For long times, the quantum system usually assumes a unique nonequilibrium steady state $\bar \rho$ that is defined by the equation $0=\sum_{\nu,j} \mathcal L_{ij}^{(\nu)}(t) \bar\rho_j(t)$ and the normalization condition $\sum_j\bar \rho_j=1$. This behavior gives rise to corresponding steady-state currents, which we denote by $J_N^{(\nu)}\equiv \underset{t\rightarrow\infty}{\lim} \dot N_\nu(t)$ and $J_E^{(\nu)}\equiv \underset{t\rightarrow\infty}{\lim} \dot E_\nu(t)$. 

Due to particle and energy conservation, these steady-state currents obey the relations $J_E\equiv J_E^{(L)}=-J_E^{(R)}$ and $J_N\equiv J_N^{(L)}=-J_N^{(R)}$.
%
%
\subsection{Equations of motion}\label{S:EoM}
%
%
In order to keep track of the time evolution of the reservoirs, we need to identify the change of their thermodynamic variables $T_\nu$, $N_\nu$, and $\mu_\nu$ with time. Neglecting the ground-state contribution in \equref{E:AverageParticleNumber}, we can derive the temperature and chemical potential changes in the reservoirs from the total differentials of the Eqs.~\eqref{E:AverageParticleNumber} and \eqref{E:InternalEnergyGeneral}, resulting in
\equ{
       \dot T_\nu &= \frac{1}{C_{\nu}} \rklamm{ {\dot U_\nu} - \frac{3}{ N_\nu \kappa_{\nu}}\dot{ N}_\nu},\label{E:TemperaturStrom}\\
       {\dot \mu_\nu} &= -\alpha_\nu \, \dot T_\nu + \frac{1}{ \kappa_{\nu} N_\nu ^2}  {\dot{ N}_\nu} \label{E:PotentialStrom}.
    }

Here, we introduce the isochoric heat capacities of the reservoirs defined as
\equ{
    C_{\nu} = \frac{\partial U_\nu}{\partial T_\nu}=  N_\nu \eklamm{12\frac{{\rm{Li}}_{4}(\xi z_\nu) }{{\rm{Li}}_{3}(\xi z_\nu) }-9\frac{ {\rm{Li}}_{3}(\xi z_\nu) }{ {\rm{Li}}_{2}(\xi z_\nu) }},
    }
the isothermal compressibility 
\equ{
    \kappa_{\nu} = \frac{1}{N_\nu^2} \left.\frac{\partial N_\nu}{\partial \mu_\nu}\right|_{T_\nu} = \frac{1}{  T_\nu N_\nu} \frac{{\rm{Li}}_{2}(\xi z_\nu)}{{\rm{Li}}_{3}(\xi z_\nu)},
    }
and the dilatation coefficients
\equ{
      \alpha_\nu =-\left.\frac{\partial  \mu_\nu}{\partial  T_\nu}\right|_{ n_\nu}=-\frac{\mu_\nu}{ T_\nu}+3\frac{{\rm{Li}}_{3}(\xi z_\nu)}{{\rm{Li}}_{2}(\xi z_\nu)}.
    }

Identifying the change of the internal energy of each reservoir $\dot U_\nu$ with the energy flow between the respective reservoir and the system, \ie, $\dot U_\nu = \dot E_{\nu}$, allows us to relate the evolution of the reservoir properties $T_\nu$ and $\mu_\nu$ to the particle and energy currents obtained from the BMS master equation. Hence, we find the relations
\begin{align}
       \begin{pmatrix}
        \dot{T}_{\nu} \\ \dot{\mu}_{\nu}
       \end{pmatrix}
       &=\frac{1}{ C_{\nu}}
       \begin{pmatrix}
           1 &  - \frac{3}{ \kappa_{\nu} N_{\nu}}	 \\
          -\alpha_\nu & \frac{3 \alpha_{\nu}}{ \kappa_{\nu} N_{\nu}} + \frac{C_{\nu}}{\kappa_{\nu} N_{\nu}^2}  
       \end{pmatrix}
       \begin{pmatrix}
	  \dot E_{\nu} \\ \dot N_{\nu}
       \end{pmatrix}\label{E:ExplicitSystem}.
\end{align}

Since the variables $T_\nu$, $N_\nu$, and $\mu_\nu$ are implicitly related via \equref{E:AverageParticleNumber}, it suffices to analyze the time evolution of two of them. However, because they are more easily accessible parameters in the experiment, it is preferable to consider the evolution of the reservoir temperatures from \equref{E:TemperaturStrom} and the evolution of the particle numbers from \equref{E:StaticCurrentDefinition} instead.
In the following, we stick to this system.

Due to the nonlinearity of the polylogarithm, we can not solve this system of coupled ordinary differential equations (ODE) analytically. Therefore, we resort to a linear-response theory which we derive in the following section.
%
%
\subsection{Linearized equations of motion}\label{S:LinearResponse}
%
%
For setups where the dimension of the system is very small compared to the dimensions of the reservoirs, the system usually runs into a quasi-steady state $\bar \rho$ on a much shorter time scale $t_{\rm{QS}}$ than the time scale of the equilibration between the reservoirs. This quasi-steady-state time scale is characterized by the rate $\Gamma_\nu(\omega)$, \ie, $t_{\rm{QS}}=1/\Gamma_\nu(\omega)$. Therefore, we can make a separation of time scales assuming the system is almost stationary during the equilibration of the reservoirs. In consequence, we are able to substitute the energy and particle currents $\dot E_{\nu}(t)$ and $\dot N_{\nu}(t)$ by their steady-state values $J_ E^{(\nu)}$ and $J_{ N}^{(\nu)}$. This fact allows us to effectively reduce the dimension of the system of coupled ODE's by considering the evolution of the temperature and particle-number differences. That leads to
\begin{align}
      \frac{\partial }{\partial t}
       \begin{pmatrix}
         {\Delta_T} \\ {\Delta_N}
       \end{pmatrix}
       &=\sum_{\nu} \frac{1}{ C_{\nu}}
       \begin{pmatrix}
           1 &  - \frac{3}{ \kappa_{\nu} N_{\nu}} \\
          0 & C_{\nu}
       \end{pmatrix}
       \begin{pmatrix}
	  J_E \\ J_N
       \end{pmatrix}\label{E:SteadyStateSystemN},
\end{align}
where we introduced the differences $\Delta_T = T_L-T_R$ and $\Delta_N = N_L-N_R$. Alternatively, we can reformulate \equref{E:SteadyStateSystemN} in terms of the linear-response steady-state heat flux $J_Q\equiv J_Q^{(L)}=-J_Q^{(R)}=J_E - (\mu+T \alpha) J_N$ \cite{Nietner2012b} corresponding to \equref{E:HeatCurrent}, which results in the equations
\begin{align}
      \frac{\partial }{\partial t}
       \begin{pmatrix}
         {\Delta_T} \\ {\Delta_N}
       \end{pmatrix}
       &=\sum_{\nu}
       \begin{pmatrix}
           \frac{1}{ C_{\nu}} &  \frac{\mu+T \alpha}{C_{\nu}} -\frac{3}{ C_{\nu} \kappa_{\nu} N_{\nu}} \\
          0 & 1
       \end{pmatrix}
       \begin{pmatrix}
	  J_Q \\ J_N
       \end{pmatrix}\label{E:SteadyStateSystemQN}.
\end{align}

Due to a linearization, here the equilibrium values of the chemical potential $\mu=[\mu_L(0)+\mu_R(0)]/2$, the temperature $T=[T_L(0)+T_R(0)]/2$, and the dilatation coefficient $\alpha$ appear explicitly.

Now, assuming that the temperature and particle-number bias between the reservoirs are symmetric about these equilibrium values, we can rewrite the reservoir temperatures and particle numbers as
\equ{
      T_L &= T+ \frac{\Delta_T}{2},\hspace{1em}T_R = T- \frac{\Delta_T}{2},\notag\\
      N_L &= N+ \frac{\Delta_N}{2},\hspace{1em}N_R = N- \frac{\Delta_N}{2}.\label{E:LinearBias}
    }

This enables us to linearize the system in \equref{E:SteadyStateSystemQN} with respect to the small differences $\Delta_T$ and $\Delta_N$, resulting in
\begin{align}
   \begin{pmatrix}
     \dot\Delta_{ T} \\ \dot\Delta_{ N}
   \end{pmatrix}
     &\approx 2
   \begin{pmatrix}
       \frac{1}{ C} &  \frac{\mu+T \alpha}{ C}-\frac{3}{C \kappa  N}	 \\
      0 &  1
   \end{pmatrix}
     {\mathcal{J}_Q}
   \begin{pmatrix}
     \Delta_{ T} \\ \Delta_{ N}
   \end{pmatrix}\label{E:LinearSystem}
   ,
\end{align}
where we defined the Jacobian matrix $\mathcal J_Q$ by
\begin{align}
  \mathcal{J}_Q = \left.
  \begin{pmatrix}
    \frac{\partial J_Q}{\partial \Delta_{ T}} & \frac{\partial J_Q}{\partial \Delta_{ N}}\\
    \frac{\partial J_N}{\partial \Delta_{ T}} & \frac{\partial J_N}{\partial \Delta_{ N}}
  \end{pmatrix} \right|_{ T,  N}\label{E:Jacobian}.
\end{align}

Note that now all reservoir properties such as heat capacity and compressibility are evaluated at the equilibrium values $T$, $N$, and $\mu=\mu(T,N)$.

The advantage of introducing the heat current $J_Q$ in favor of the energy current $J_ E$  in the above equations is, that the Jacobian $\mathcal{J}_Q$ can now be related to linear-response transport coefficients in correspondence with our previous work \cite{Nietner2013}.

Introducing the definition of the positive particle conductivity as
\equ{
      \sigma \equiv -\frac{J_N}{\Delta_\mu} \approxeq -\kappa  N^2 \frac{ J_N}{ \Delta_{ N}},\hspace{.5em}{\rm{for}}\hspace{.5em}\Delta_T=0,\label{E:sigmaDefinition}
    }
the definition of the positive heat conductivity as
\equ{
      q \equiv - \frac{J_Q}{ \Delta_{ T}}, \hspace{.5em}\textrm{for}\hspace{.5 em}J_N=0,\label{E:heatDefinition}
    }
and the definition of the Seebeck coefficient at vanishing particle current as
\equ{
      \Sigma \equiv -\frac{\Delta_\mu}{\Delta_T} \approxeq -\frac{1}{ \kappa  N^2}\frac{ \Delta_{ N}}{ \Delta_{ T}} \label{E:SeebeckDefinition},\hspace{.5em}\textrm{for}\hspace{.5em}J_N=0,
    }
we find that the Jacobian from \equref{E:Jacobian} can be reformulated in terms of linear-response transport coefficients which yields
\begin{align}
  \mathcal{J}_Q =
  \begin{pmatrix}
    -q-T\sigma \Sigma^2& T\frac{\sigma\Sigma}{ \kappa  N ^2}\\
    \sigma \Sigma & -\frac{\sigma}{ \kappa  N^2}
  \end{pmatrix}.
\end{align}

Here, we additionally used the Onsager reciprocal relation $T \,{\partial J_N}/{\partial\Delta_T}=N^2 \kappa\, {\partial J_Q}/{\partial\Delta_N}$ \cite{Onsager1931a,Onsager1931b}. 

Since all matrix elements are evaluated in the equilibrium, and hence are time independent, we can solve the system in \equref{E:LinearSystem} exactly resulting in
\begin{widetext}
\equ{
      \Delta_T(t) &= \left[\left(\frac{\sigma }{N^2\kappa }+\frac{\Omega -\delta }{2}\right)\Delta_T{(0)} -\frac{\sigma \mu_{\rm{eff}} }{N^2 \kappa C} \Delta_N{(0)} \right] \frac{e^{t(\Omega -\delta )}}{\Omega }\left[ 1-\frac{\left(\frac{\sigma }{N^2\kappa }-\frac{ \Omega +\delta }{2}\right)\Delta_T{(0)} -\frac{\sigma \mu_{\rm{eff}} }{N^2 \kappa C} \Delta_N{(0)}}{\left(\frac{\sigma }{N^2\kappa }+\frac{\Omega-\delta }{2}\right)\Delta_T{(0)} -\frac{\sigma \mu_{\rm{eff}} }{N^2 \kappa C} \Delta_N{(0)}} e^{-2\Omega  t}\right],\label{E:FullEoMSolutionT}\\
      \Delta_N(t) &= \left[\sigma  \Sigma  \Delta_T{(0)}-\left(\frac{\sigma }{N^2\kappa }-\frac{\Omega +\delta}{2}\right)\Delta_N{(0)}\right] \frac{e^{t(\Omega -\delta )}}{\Omega } \left[ 1-\frac{\sigma  \Sigma  \Delta_T{(0)}-\left(\frac{\sigma }{N^2\kappa }+\frac{\Omega-\delta }{2}\right)\Delta_N{(0)}}{\sigma  \Sigma  \Delta_T{(0)}-\left(\frac{\sigma }{N^2\kappa }-\frac{\Omega+\delta }{2}\right)\Delta_N{(0)}} e^{-2\Omega  t}\right],\label{E:FullEoMSolutionN}
    }
\end{widetext}
with the effective chemical potential
\equ{
    \mu_{\rm{eff}} = \mu + T \alpha - \frac{3}{N \kappa },
    }
that reflects the modifications arising from the presence of a temperature- and particle-number bias. 

Furthermore, we introduced the positive coefficient
\equ{
      \delta = \frac{\sigma}{N^2 \kappa } + \frac{q+T \sigma \Sigma ^2}{C} - \frac{\sigma \Sigma }{C}\mu_{\rm{eff}},
    }
and the positive real frequency 
\equ{
      \Omega =\sqrt{ \delta^2 - \frac{4}{C}\det[\mathcal{J}_Q]} =\sqrt{ \delta^2 - \frac{4 }{C} \frac{\sigma q}{N^2 \kappa} }.
    }

The properties of the few-level quantum system enter in these expressions via the linear-response transport coefficients $\sigma$, $\Sigma$, and $q$.

In general, the determinant of the Jacobian $\mathcal J_Q$ is nonvanishing and positive, \ie, $\det[\mathcal{J}_Q] \ge 0$, which leads to the fact that one always finds
  $\Omega\leq\delta.$
Here, the equality only occurs in the limit where the energy and particle current are tightly coupled, \ie, $\dot E_\nu = \omega \dot N _\nu$. This proportionality results in  a vanishing heat conductivity, \ie, $q=0$, and hence in a vanishing determinant $\det[\mathcal{J}_Q]= 0$.

Taking a look at the linear evolutions in \equref{E:FullEoMSolutionN}, we note that they consist of the product of two exponential processes. First, we have a saturation process that is characterized by the time scale 
\begin{equation}
  t_<=\frac{1}{2\Omega}.\label{E:SmallTimeScales}
\end{equation}

This process dominates for short times. Using the initial condition $\Delta_N(0)=0$ and $\Delta_T(0)\neq 0$, we find that this process leads to an initial increase of the particle-number bias up to a maximum value. This maximum is reached at time $t_{\rm{max}}$, which explicitly reads as
  $t_{\rm{max}}={1}/{(2\Omega)}\ln\left[(\delta+\Omega)/(\delta-\Omega)\right].$
For longer times, the evolutions are dominated by an exponential decay process, which is characterized by the time scale 
\begin{equation}
  t_> = \frac{1}{\delta-\Omega}.\label{E:LongTimeScales}
\end{equation}

Note that the latter time scale is not defined in the tight-coupling limit. In fact, in this limit there is no exponential decay and, hence, the thermodynamic reservoir variables remain maximally biased, as exemplarily shown in \figref{fig::FerSingleLev}.
%
%
\subsection{Efficiency of a Heat Engine}\label{S:Efficiency}
%
%
Finally, we note that in a thermodynamic device as shown in \figref{F:GeneralTransportScheme}, the initial nonequilibrium configuration can be used to perform work. 

In order to analyze the efficiency with which power can be extracted from the device, we use the Shannon entropy of the system given by $S=-\tr{\rho \ln \rho}$. Performing a differentiation with respect to time, one obtains the change of the Shannon entropy as 
\equ{
    \dot S = - \sum_i \dot \rho_i \ln \rho_i,\label{E:EntropyProduction} 
    } 
where the sum runs over all energy eigenstates of the system Hamiltonian $\hat{\mathcal{H}}_{\rm {S}}$.

Using the master equation from \equref{eq::rhofermi}, we can calculate this 
entropy production. We observe that it can be decomposed into the sum $\dot S =\dot S_i +\dot S_e $ of the internal entropy production $\dot S_i\ge 0$ and the entropy flow from the reservoirs $\dot S_e=\sum_\nu \beta_\nu \dot Q_\nu$. In the quasi-steady-state regime, the change of the Shannon is approximately zero, such that the entropy production can be written as \cite{Esposito2010a}
\equ{
    \dot S_i \approxeq -\sum_\nu \beta_\nu \dot Q_\nu \ge 0.
    }

From this entropy production, we can derive a bounded efficiency measure \cite{Broeck2005}. We are especially interested in the conversion of the heat current 
\equ{
    \dot Q_{\rm in}\equiv\dot E_{\rm hot} - \mu_{\rm hot} \dot N_{\rm hot}\label{E:Qin},
    }
flowing from the hot reservoir with temperature $T_{\rm hot}$ to the cold reservoir with temperature $T_{\rm cold}$ into power $P$ that can be extracted from the device. Therefore, we define the instantaneous efficiency as
\equ{
    \eta_{}(t) = \frac{P(t)}{\dot Q_{\rm in}(t)}\le\eta_C(t)\label{E:Eta},
    }
where we introduced the instantaneous Carnot efficiency as $\eta_C(t)=1-{T_{\rm cold}(t)}/{T_{\rm hot}(t)}$ \cite{Broeck2012}. However, this should not be confused with cyclic efficiencies, since here we are just considering the efficiency of a single overall relaxation process. 


Noticing that the device performs chemical work by shifting particles against a chemical bias, we find that the power output of the device in the form of chemical work is defined by 
\equ{
    P(t)\equiv-\dot W(t)= -\sum_\nu \mu_\nu (t)\dot N_\nu (t)\label{E:Work}.
    }

The chemical work is defined such that it is negative, if the system performs work, and positive, if work is performed on the system. Hence, we are solely interested in the work mode where one obtains a positive power output $P(t)\ge 0$. 

Additionally to the instantaneous efficiency in \equref{E:Eta}, we can also integrate this quantity to yield the cumulative efficiency
\equ{
    \eta_{\rm cum}(t) \equiv \frac{\int_{0}^{t} P(t^\prime)dt^\prime}{\int_{0}^{t}\dot Q_{\rm in}(t^\prime)dt^\prime}=\frac{-W(t)}{Q_{\rm in}(t)}\label{E:EtaTot},
    }
which is given by the ratio of the total work $-W$ performed on the reservoirs and the consumed heat $Q_{\rm in}$.

In a recent experimental setup, the cumulative efficiency from \equref{E:EtaTot} has also been measured for two ultracold atomic reservoirs connected via a narrow 2D channel  \cite{Brantut2013}.
%
%
\section{Results}\label{S:Results}
%
%
From our numerical simulations of the full set of coupled ODE's from Eqs.~\eqref{eq::rhofermi}--\eqref{E:StaticEnergyCurrentDefinition}, we find that, depending on the number of allowed transition energies in the quantum system, one can distinguish two qualitatively different cases for the evolution of the full system. Namely, the single-transition-energy, \ie, tight coupling, and the multi-transition-energy case. Since each transition energy of the system opens up a corresponding channel which allows for particle transport, we also refer to these cases as the single and multi-transport-channel situations. We elaborate these different cases in more detail within the subsequent paragraphs. 
%
%
\subsection{Single Fermionic Transport-Channel}\label{sec::fersingle}
%
%
The Hamiltonian for a fermionic system with a single transition energy $\omega$ reads as
\equ{
    \mathcal{H}_S= \omega \hat{a}^\dag\hat{a}.
    }

This system has only two different states, the vacuum state with the population $\rho_0$ and energy $0$, and the single-particle state corresponding to the population $\rho_1$ with energy $\omega$, that equals the transition energy from $0$ to $1$ particle in the system. For such a system, we find that the respective particle and energy currents defined in Eqs.~\eqref{E:StaticCurrentDefinition} and \eqref{E:StaticEnergyCurrentDefinition} become (see Appendix \ref{A:Liouvillian})
\equ{
    \dot{N}_\nu(t) = \Gamma_\nu(\omega) \left[\rho_1-\bar n_\nu^{(-)}(\omega)\right],\hspace{1em}\dot{E}_\nu(t)=\omega \dot{N}_\nu(t). \label{E:FermiSingelLevelCurrents}
}

Thus, in the case of a single transition energy in the quantum system, we obtain the tight-coupling limit.
In this limit, the transport of heat through the quantum system at vanishing particle current is not possible and, therefore, a full equilibration of the reservoirs can not be achieved. 
\begin{figure}[t]
  \begin{minipage}[t]{.95\columnwidth}
   \begin{overpic}[width=.95\columnwidth]{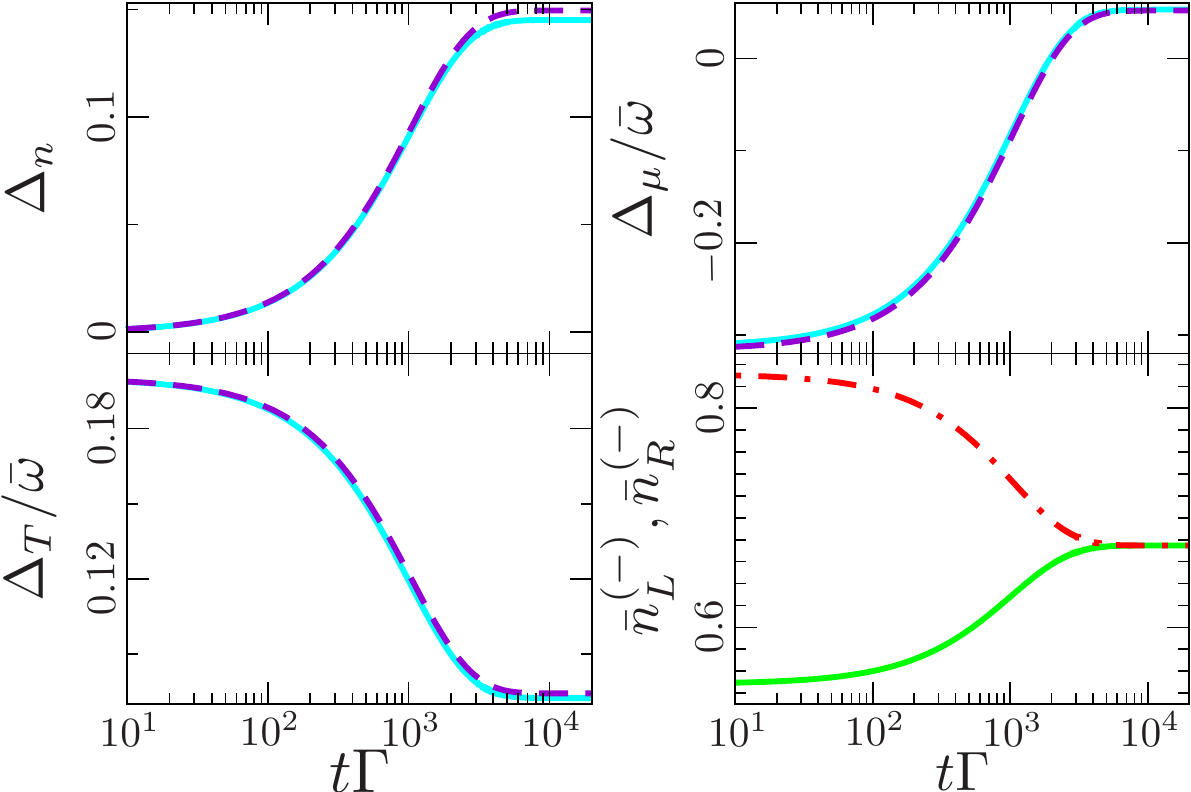}
      \put(13,60){(a)}
      \put(65,60){(b)}
      \put(13,28){(c)}
      \put(65,28){(d)}
   \end{overpic}
   \caption{
	   (Color online) Plot of the time evolution of the differences $\Delta_n = (N_L-N_R)/\mathcal N$, $\Delta_T=T_L-T_R$, and $\Delta_\mu=\mu_L-\mu_R$, of the thermodynamic variables of the left and right reservoirs, for a quantum system with a single transition energy $\omega=0.9\bar{\omega}$. The solid lines correspond to the numeric simulation, and the dashed lines to the linear-response solution from Eqs.~\eqref{E:FullEoMSolutionT} and \eqref{E:FullEoMSolutionN}. The initial particle numbers are set to $N_L(0)=N_R(0)=0.5\mathcal N$ and the temperatures to $T_L(0)=0.45\bar{\omega}$ and $T_R(0)=0.25\bar{\omega}$. In panel (d), we show the evolution of the Fermi functions of the right (dotted-dashed line) and left (solid line) reservoirs.
	   }
   \label{fig::FerSingleLev}
  \end{minipage}
\end{figure}
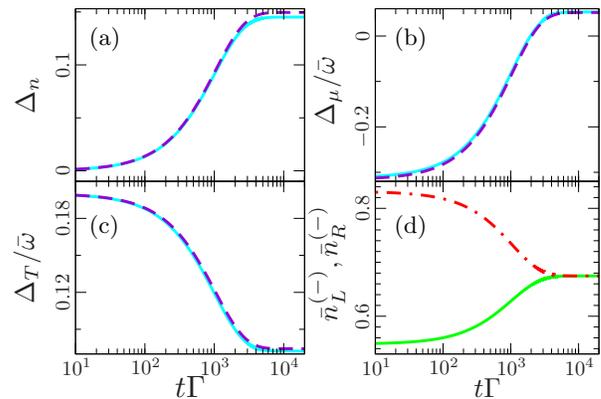

This behavior is confirmed by our numerical simulation shown in \figref{fig::FerSingleLev}. Here, we assume that initially the quantum system is empty and the two reservoirs are in a nonequilibrium configuration with the same particle densities but different temperatures. Subsequently, we let the full system evolve in time until it reaches its steady state. This steady state is achieved when the net currents through the quantum system vanish, \ie, when the Fermi functions of the reservoirs evaluated at the system transition energy $\omega$ are the same. We find that the transport process through this quantum system reaches a steady state, however, this state is not a thermal equilibrium state, since the thermodynamic variables $T_\nu$, $N_\nu$, $\mu_\nu$ of the left and right reservoirs differ. 

\begin{figure}[t]
  \begin{minipage}[t]{.95\columnwidth}
   \begin{overpic}[width=.95\columnwidth]{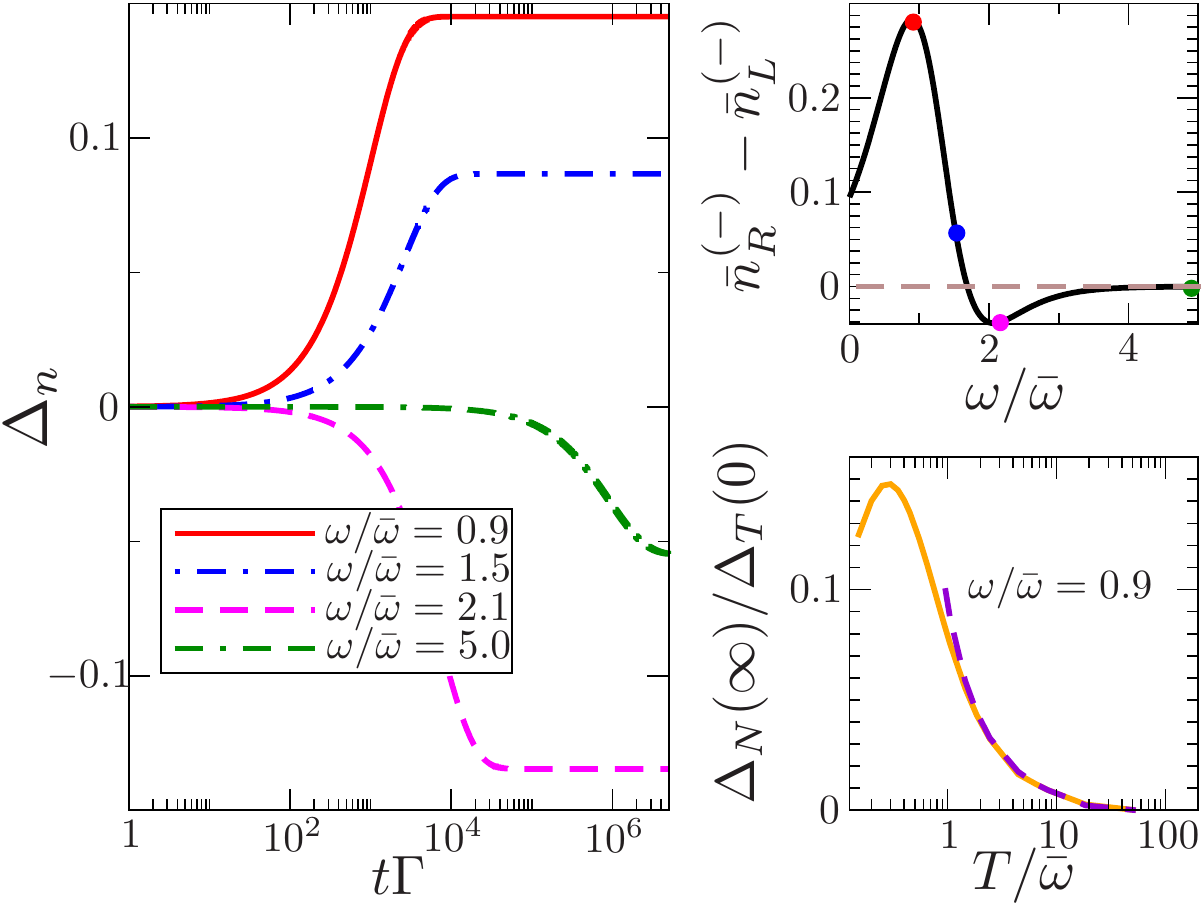}
      \put(13,68){(a)}
      \put(92,68){(b)}
      \put(92,31){(c)}
   \end{overpic}
   \caption{(Color online) 
	   (a) Time evolution of the particle-number difference $\Delta_ n = (N_L-N_R)/\mathcal N$ for different values of the quantum system transition energy $\omega$ for the same initial parameters as in Fig.~\ref{fig::FerSingleLev}. %
	   (b) Plot of the difference $\bar n_L^{(-)}(\omega)-\bar n_R^{(-)}(\omega)$ of the initial Fermi functions versus the transition energy of the system. The dots correspond to the energies $\omega$ in panel (a). %
	   (c) Plot of the ratio between the steady-state particle-number bias and the initial temperature bias $\Delta_T(0) =0.2\bar{\omega}$ in dependence of the average temperature $T$ of the system, for a fixed transition energy $\omega=0.9\bar{\omega}$. The $y$ axis is measured in units $1/\bar\omega$. For high temperatures, the particle bias vanishes like $1/T$ for, both, fermions (solid line) and bosons (dashed line).
	   }
  \label{fig::FerSingleDirCurrent}
  \end{minipage}
\end{figure}

In particular, we find that for an initial temperature bias and equal particle numbers, a finite particle-number difference builds up, as the system evolves. This effect is accompanied by a decrease of the initial thermal bias and chemical potential bias. The amount of these differences in the thermodynamic reservoir properties, and therefore also the sign of the resulting bias, can be tuned by changing the system transition energy. We demonstrate this effect in \figref{fig::FerSingleDirCurrent}(b), where we plot the difference of the Fermi functions $\bar n_\nu^{(-)}(\omega)$ of the left and right reservoir for several values of the system transition energy $\omega$.

Here, we observe either an increase or a decrease of the particle number in the reservoir, depending on the system transition energy in relation to the threshold energy $\omega_0$. 
%
%
This threshold energy is defined be the equality of the Fermi functions of the left and right reservoirs at constant chemical potentials and constant temperatures, \ie, $\bar n_L^{(-)}(\omega_0)=\bar n_R^{(-)}(\omega_0)$, which corresponds to
\equ{
    \omega_0\equiv \frac{T_L\mu_R-T_R\mu_L}{T_L-T_R}
    \label{eq::direction}.
    }

If the system transition energy  lies above the threshold, \ie, $\omega>\omega_0$, the particle current flows from right to left, \ie, with the chemical potential bias. Otherwise, one observes a flow against the chemical potential bias. Note, that also the velocity of the change in particle number is altered.

In order to quantify this effect, we analyze the single-transport-channel setup using the linear approach introduced in \secref{S:LinearResponse}. From \equref{E:FullEoMSolutionN} we see that the steady-state particle-number bias $\Delta_N(\infty) \equiv \lim_{t\rightarrow\infty}\Delta_N(t)$ is given by 
\begin{align}
 \Delta_N(\infty) = \frac{\sigma  \Sigma}{\Omega } \Delta_T(0)  = \frac{C (\mu+\alpha T-\omega) \Delta_T(0)}{\frac{C T}{N^2 \kappa } - (\mu+\alpha T-\omega)(\omega-\frac{3}{N \kappa})} ,\label{E:SteadStateDensityBias}
\end{align}
where we used the tight-coupling limit result
\equ{
    \Omega=\delta=\frac{\sigma}{N^2 \kappa } + \frac{T \sigma \Sigma ^2}{C} - \frac{\sigma \Sigma }{C}\mu_{\rm{eff}}.
    }

The respective linear-response transport coefficients have been calculated from the steady-state particle and energy currents through the system (see Appendix \ref{A:FermiOneLevelCoefficients}). 

From \equref{E:SteadStateDensityBias} we see that the steady-state particle-number bias $\Delta_N(\infty)$ approaches a finite maximum value as the system transition energy  approaches zero.
Contrary, if the transition energy is increased, the steady-state particle number vanishes eventually, \ie, $\lim_{\omega\rightarrow \infty}\Delta_N(\infty)=0$, in correspondence with Figs.~\ref{fig::FerSingleDirCurrent}(a) and \ref{fig::FerSingleDirCurrent}(b). In between, there is a finite transition energy $\omega = \mu+\alpha T$, where the steady-state particle number also vanishes. Here, the energy of the transport channel is exactly equal to the chemical potential $\mu$ in the reservoirs plus the contribution to the chemical potential $\alpha T$ arising from the presence of a small temperature bias. Note that this energy value is the linear response equivalent to the threshold energy form \equref{eq::direction}.
\begin{figure}[t]
  \begin{minipage}[t]{.95\columnwidth}
   \begin{overpic}[width=.95\columnwidth]{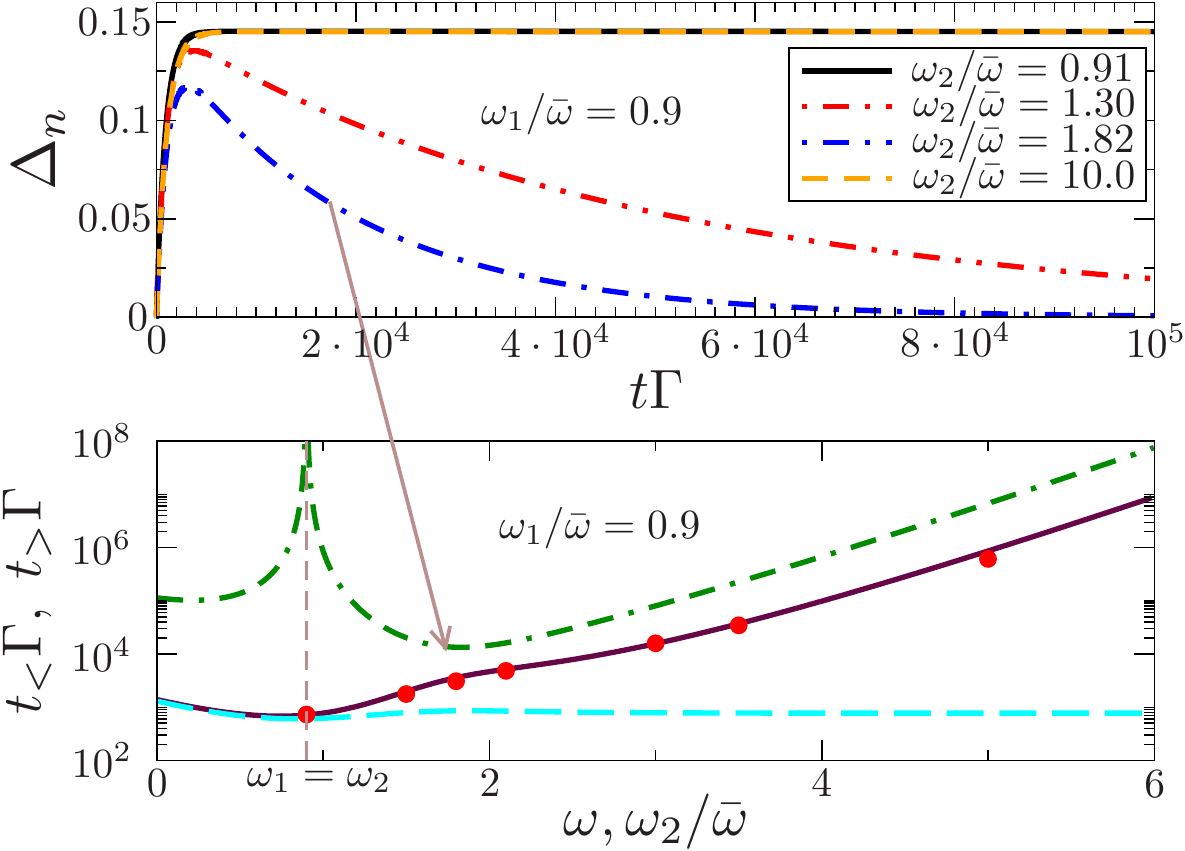}
      \put(-1,68){(a)}
      \put(-1,34){(b)}
   \end{overpic}
  %
  \caption{(Color online) 
	  (a) Plot of the time-evolution of the particle-number difference $\Delta_n=(N_L-N_R)/\mathcal N$ for a system with two different transition energies. %
	  (b) Plot of the linear-response results for the characteristic time scales $t_<$ from \equref{E:SmallTimeScales} and $t_>$ from  \equref{E:LongTimeScales} for a system with one (solid line) and two (dashed, dotted-dashed lines) transition energies in dependence of one of the energies. In the case of one channel, we compare it with the full numerics (dots). For the case of two transport channels, the lower transition energy is fixed to $\omega_1=0.9\bar \omega$ in both plots. The arrow indicates the fastest thermalization process corresponding to the minimum of $t_>$. 
	  }
  \label{fig::Ener_diff}
  \end{minipage}
\end{figure}

Moreover, we note that this nonequilibrium steady-state results from the discrete energy structure of the few-level quantum system and thus is a pure quantum mechanical effect. Consequently, we observe in \figref{fig::FerSingleDirCurrent}(c), that for high temperatures the finite steady-state particle-number bias vanishes leading to the classically expected result of equilibrated reservoirs.

With these results, we can also analyze the characteristic time scale $t_<$ which is shown in \figref{fig::Ener_diff} (solid line). Here, we find that this time scale increases exponentially with the system transition energy $\omega$. This effect is caused by the particle conductance which exponentially decreases with increasing energy, because for increasing energy, the corresponding occupations in the reservoirs become exponentially small. 
\begin{figure}[t]
  \begin{minipage}[t]{.95\columnwidth}
   \begin{overpic}[width=.95\columnwidth]{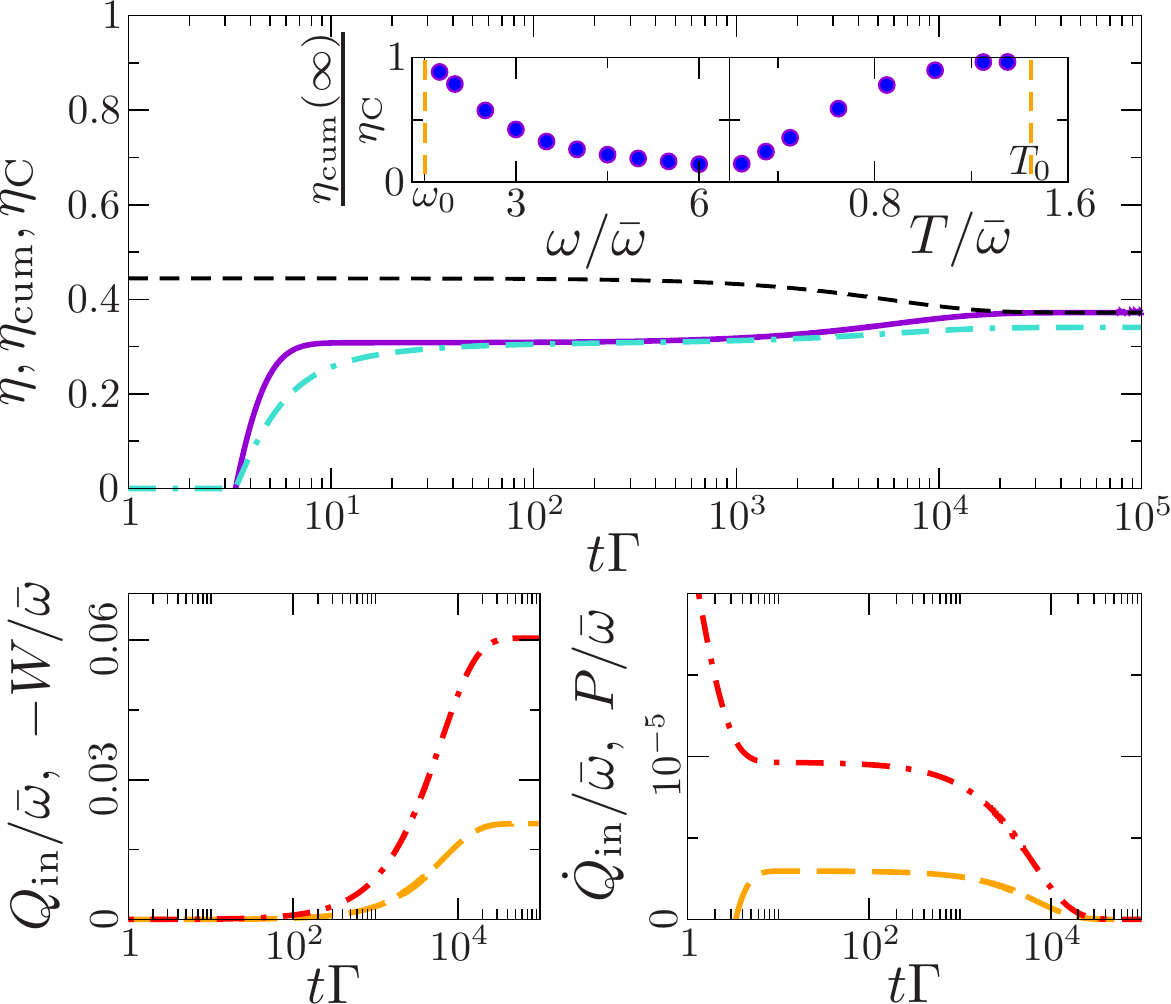}
      \put(13,78){(a)}
      \put(90,29){(c)}
      \put(13,29){(b)}
   \end{overpic}
   \caption{(Color online) 
       Plots of the numerically obtained efficiency for a single transport channel with $\omega/\bar \omega= 2$. The initial parameters are set to $N_L(0)=N_R(0)=0.5\mathcal N$, $T_L(0)=0.45\bar{\omega}$, and $T_R(0)=0.25\bar{\omega}$. %
       (a) Plot of the time evolution of the instantaneous efficiency from \equref{E:Eta} (solid line), the cumulative efficiency from \equref{E:EtaTot} (dotted-dashed line), and the instantaneous Carnot efficiency (dashed line). The insets show the steady-state total efficiency (dots) for different energies of the system and for different average temperatures $T=(T_L+T_R)/2$ of the reservoirs. Here, a cut-off appears (dashed line) beyond which no power can be extracted from the device. %
       In panel (b) we show the corresponding total heat (dotted-dashed line) and work (dashed line) and %
       in panel (c) the corresponding  instantaneous heat flow from \equref{E:Qin} (dotted-dashed line) and power from \equref{E:Work} (dashed line).
	   }
   \label{fig::EtaFerSingleLev}
  \end{minipage}
\end{figure}

Additionally, we investigate the efficiency of the process that converts a thermal bias into a particle-number bias in \figref{fig::EtaFerSingleLev}. To this end, we numerically calculate the heat flow into the system $\dot Q_{\rm in}(t)$ and the power $P(t)$ extracted from the device according to the \equref{E:Work} and insert them into the definitions in \equref{E:Eta} and \equref{E:EtaTot}. In \figref{fig::EtaFerSingleLev} we show some of the results. 

First, we notice that the power output of the device is not always positive. For example, this can be observed in  \figref{fig::EtaFerSingleLev}(a). In the transient regime of small times when the system gets filled, work has to be done on the system and the corresponding efficiency is set to zero in this regime. This is due to the fact, that the efficiency is only defined for a positive power output. For larger times, the system enters the quasi-steady-state regime where we observe a finite efficiency of about $\eta \approx 0.6 \eta_C$. When the whole system enters its steady state, the instantaneous efficiency (solid line) trivially becomes maximal, \ie, $\eta=\eta_C$, since all currents vanish. We also show the cumulative efficiency (dotted-dashed line) which assumes a finite steady-state value $\eta_{\rm cum}(\infty)\approx 0.9 \eta_C$. 

This finite steady-state efficiency is further investigated in the insets of \figref{fig::EtaFerSingleLev}(a). Here, we find that quite high efficiencies can be achieved in this setup, depending on the transition energy $\omega$ and the average temperature $T$. The efficiency is increased for small transition energies and large temperatures. However, for these parameters the steady-state particle-number bias is also diminished. Hence, a careful tuning of these parameters with respect to an optimal efficiency to bias ratio is necessary. 

Furthermore, we find that there are threshold values beyond which the system does not perform work (dashed lines). For low temperatures $T<T_0$ and a fixed transition energy $\omega$ in the system, we observe a flow of particles against the chemical potential bias. Therefore, the device performs chemical work.
Contrary, for high temperatures $T>T_0$, we observe a flow of particles with the chemical potential bias and, hence, work is done on the device and the efficiency to extract power is not defined. In correspondence with the energy threshold from \equref{eq::direction}, the temperature threshold is defined by the equality of the left and right reservoir Fermi functions for a given energy $\omega$ and constant chemical potentials, \ie, $\bar n_L^{(-)}(\omega)=\bar n_R^{(-)}(\omega)$. The resulting expression reads as
\equ{
    T_0=\frac{T_L\rklamm{\mu_L+\mu_R-2\omega}}{2(\mu_L-\omega )},
    }
where the temperature threshold $T_0$ is the average temperature, \ie, $T_0\equiv{T_L+T_R}/{2}$.

An analogous argument holds when we consider constant reservoir temperatures and vary the system transition energy. For decreasing energy, the efficiency increases. However, below the transition energy threshold from \equref{eq::direction} no power can be extracted from the device.

Additionally, we show the overall performed work in comparison to the total heat in \figref{fig::EtaFerSingleLev}(b). Moreover, in order to identify the time domain of maximum power output, we show the instantaneous power and heat current in \figref{fig::EtaFerSingleLev}(c). We observe that the power output is maximal in the quasi-steady-state regime, whereas it is almost zero for very small and very large times.
%
%
\subsection{Multiple Fermionic Transport-Channels}\label{sec::fertwo}
%
%
As an example for a transport setup with multiple transport channels, we now consider the situation of a fermionic quantum system with two internal transition energies. To this end, we model the quantum system by the Hamiltonian 
\equ{
    \mathcal{H}_S=  \sum_{i=1}^2 \omega_i \hat{a}_i^\dag\hat{a}_i .
    }

Here, we assume that only two transitions are possible in the system, namely, the transition from the vacuum state $\rho_0$ to the one-particle state $\rho_1$ with energy $\omega_1$, and the transition from the vacuum to the one-particle state $\rho_2$ with energy $\omega_2$. These two transition energies give rise to two possible transport channels which contribute to the overall energy and particle currents through the system. These currents read explicitly (see Appendix \ref{A:Liouvillian})
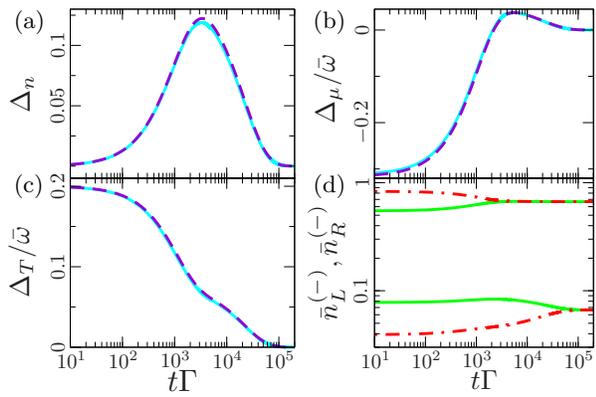
\begin{figure}[t]
  \begin{minipage}[t]{.95\columnwidth}
   \begin{overpic}[width=.95\columnwidth]{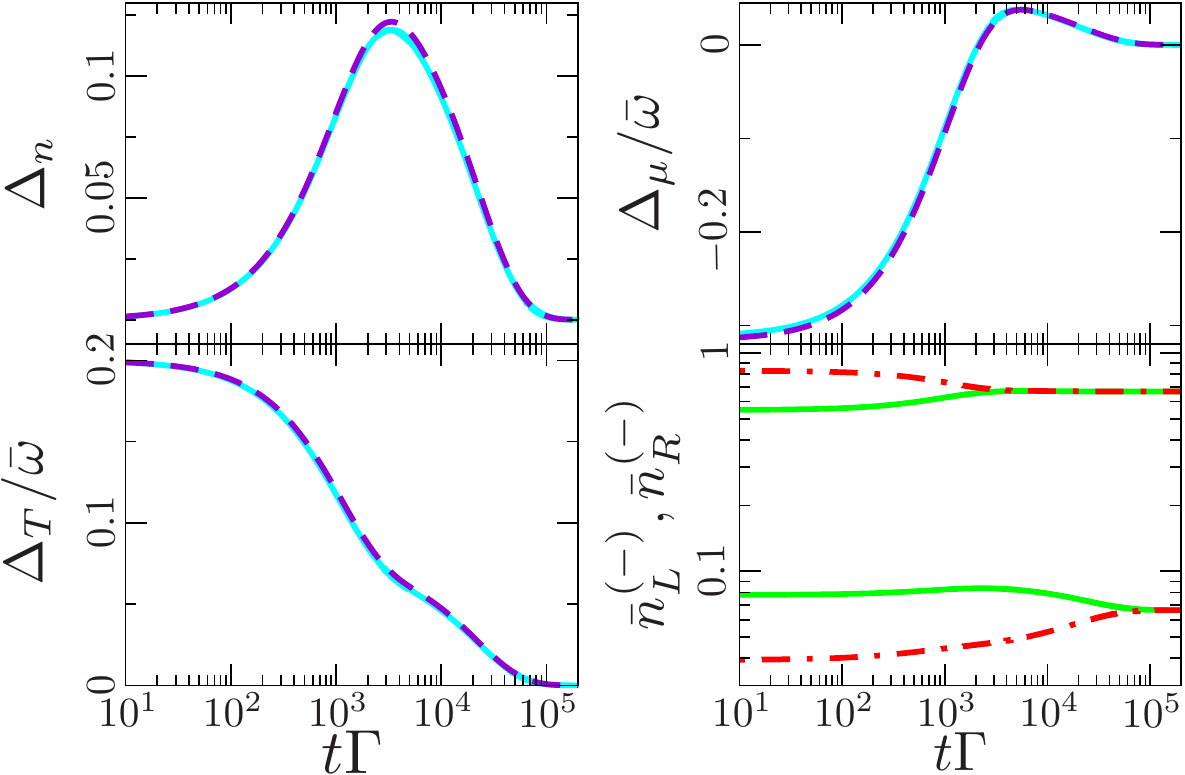}
      \put(1,62){(a)}
      \put(52,62){(b)}
      \put(1,34){(c)}
      \put(52,34){(d)}
   \end{overpic}
  \caption{(Color online) 
	  Plot of the time-evolution of the differences $\Delta_n = (N_L-N_R)/\mathcal N$, $\Delta_T=T_L-T_R$ and $\Delta_\mu=\mu_L-\mu_R$, of the thermodynamic variables of the left and right reservoirs, for a quantum system with two transition energies $\omega_1=0.9\bar{\omega}$ and $\omega_2=2.1\bar{\omega}$. The solid lines correspond to the numeric simulation, and the dashed lines to the linear-response solution. The initial particle numbers are set to $n_L(0)=n_R(0)=0.5$ and the temperatures to $T_L(0)=0.45\bar{\omega}$ and $T_R(0)=0.25\bar{\omega}$. %
	  In panel (d), we plot the Fermi functions of the right (dotted-dashed line) an left (solid line) reservoir for both transport channels. The upper branch corresponds to the transition energy $\omega_1$ and the lower branch to $\omega_2$.
	  }
  \label{fig::FerTwoLev}
  \end{minipage}
\end{figure}
\begin{align}
  \dot{E}_\nu(t)=&\sum_{i=1}^2 \omega_i \Gamma_\nu(\omega_i) [\rho_i - \bar n^{(-)}_\nu (\omega_i)(\rho_i+\rho_0)],\label{E:FermiTwoLevelEnergyCurrent}\\
  \dot{N}_\nu(t)=&\sum_{i=1}^2 \Gamma_\nu(\omega_i) [\rho_i - \bar n^{(-)}_\nu (\omega_i)(\rho_i+\rho_0)].\label{E:FermiTwoLevelCurrent}
\end{align}

Together with the corresponding equations for the evolution of the system density matrix from \equref{eq::rhofermi}, these currents determine the equilibration process between the attached reservoirs. In \figref{fig::FerTwoLev}, we present some numerical results for the thermodynamical variables of the reservoirs. 

Initializing the reservoirs with equal particle numbers and a finite-temperature bias $\Delta_T(0)\neq 0$ between them, we observe the buildup of a particle-number bias. This bias reaches a maximum where the overall particle current vanishes. However, contrary to the case with a single transport channel, the energy and particle currents vanish independently. Consequently, the finite-energy current at vanishing particle current allows the system to further relax and the steady state is reached only when all thermodynamic variables are in equilibrium. 

In order to gain some analytic insight, we also calculate the steady-state currents (see Appendix \ref{A:FermiTwoLevelCoefficients}) corresponding to the expressions in \equref{E:FermiTwoLevelCurrent}. Subsequently, we extract the respective linear-response transport coefficients, which are inserted into the Eqs.~\eqref{E:FullEoMSolutionN}. In comparison, we find a good accordance of the linear-response theory to the numerical solution.

From our simulations, we further observe that, for the two-transport-channel system, we can control the dynamics by tuning the difference $\Delta_\omega=\omega_2-\omega_1$ between the two transition energies of the quantum system. In Fig.~\ref{fig::Ener_diff}(a), we show the evolution of the particle-number bias for different values of $\Delta_\omega$. We find that, for very large, as well as for very small differences, the evolution of the reservoir particle number resembles the single-transport-channel result, leading to a very slow equilibration of the reservoirs. In between, we observe a regime where the reservoirs equilibrate considerably faster.

This effect is further illustrated in Fig.~\ref{fig::Ener_diff}(b), where we plot the time scales $t_<$ from  \equref{E:SmallTimeScales} (dashed line) and $t_>$ from \equref{E:LongTimeScales} (dotted-dashed line) for different values of $\Delta_\omega$, assuming a fixed frequency $\omega_1$. Here, we see that the first time scale $t_<$ is hardly affected by the difference of the transition energies. Apart from a small decrease around $\Delta_\omega=0$, this time scale stays almost constant around the value $t_< \Gamma\approx 10^3$). The reason for that behavior can be traced back to the fact that this time scale is mostly affected by the particle conductance $\sigma$, which itself is largely influenced by the lowest transition energy, whereas the higher transition energies have an exponentially suppressed contribution to the particle current.

Contrary, the time scale $t_>$, which characterizes the exponential decay into the thermodynamic equilibrium, strongly depends on the difference of the transition energies. It diverges when approaching the single-transport-channel configuration where $\Delta_\omega=0$, and grows exponentially for large energy differences. This diverging behavior can be traced back to the fact that this time scale is mostly affected by the heat conductance $q$, which explicitly depends on the difference of the transition energies.

Here, we point out that this effect can be used to define a particle transistor or particle capacitor for ultra cold gases. By shifting the upper transition energies to high values with respect to the lowest transition energy, one can use, \eg, an initial temperature bias to establish a particle-number bias between the reservoirs. This difference can be maintained within the setup for very long times. However, by lowering the energy of the upper transition energies of the system, the equilibration between the reservoirs can be triggered, leading to a controlled decrease of the reservoir bias, and eventually to a full equilibration.
%
%
\subsection{Multiple Bosonic Transport-Channels}\label{sec::bosetwo}
%
%
For completeness, we now briefly consider a bosonic transport system with two transport channels. This system underlines the fact that the method proposed within this paper is also applicable on interacting system.

In the following, we consider two reservoirs of massive bosonic particles confined in an inhomogeneous harmonic trap, which are governed by the Hamiltonian from \equref{E:BathHamiltonian}. These reservoirs are weakly coupled to a quantum system which is described by the Hamiltonian
\begin{align}
  \hat{\mathcal H}_S=\varepsilon\hat{a}^\dag\hat{a}+\frac{\phi}{2}\hat{a}^\dag\hat{a}(\hat{a}^\dag\hat{a}-1),\label{E:BoseHamiltonian}
\end{align}
where $\phi$ is two-particle interaction energy. The bosonic operators $\hat a$ and $\hat a ^\dagger$ annihilate and create a particle with energy $\varepsilon$, respectively. Consequently, the energy spectrum is given by $\omega_n= n \varepsilon +\phi n (n-1)/2$, where $n$ corresponds to the number of particles in the system. In order to restrict this system to two transition energies only, we truncate the system Hilbert space at $n=2$.

\begin{figure}[t]
  \begin{minipage}[t]{.95\columnwidth}
   \begin{overpic}[width=.95\columnwidth]{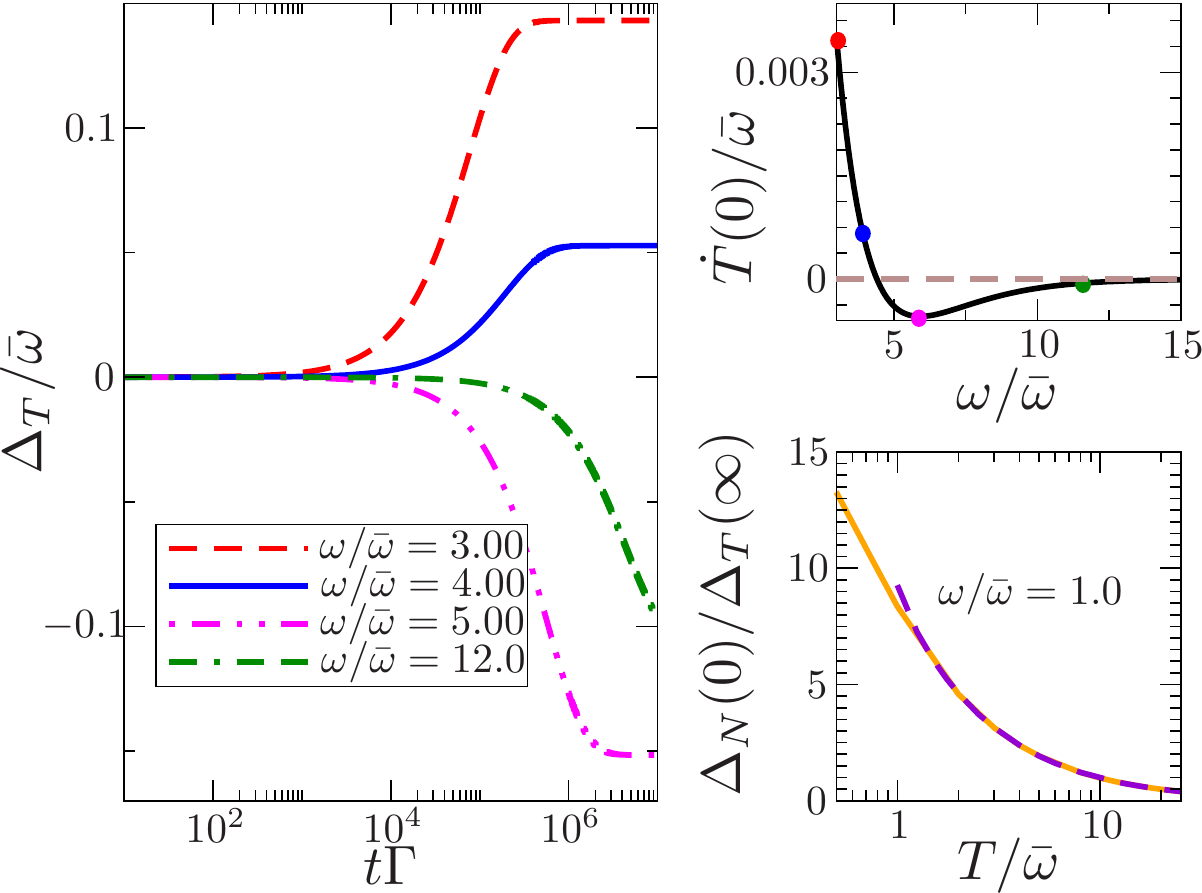}
      \put(12,68){(a)}
      \put(90,68){(b)}
      \put(90,30){(c)}
   \end{overpic}
   \caption{(Color online) 
	   (a) Time evolution of the temperature difference $\Delta_T$ in a bosonic transport setup, for different values of the system transition energy $\omega$ with the initial particle numbers $N_L(0)=0.6\mathcal N$ and $N_R(0)=0.4\mathcal N$ and the fixed average temperature $T=1.5\bar \omega$. %
	   (b) Plot of the initial temperature current $\dot T_L(0)$ versus the transition energy of the system. The dots correspond to the energies $\omega$ in panel (a). %
	   (c) Plot of the ratio between the initial particle-number bias and the steady-state temperature bias $\Delta_T(0) =0.2\bar{\omega}$ in dependence of the average temperature $T$ of the system, for a fixed transition energy. The $y$ axis is measured in units $1/\bar\omega$. For high temperatures, this ratio vanishes like $1/T$ for both, fermions (solid line) and bosons (dashed line).
	   }
  \label{fig::BoseSingleDirCurrent}
  \end{minipage}
\end{figure}

Subsequently, the energy and particle currents through the quantum system are obtained from the respective Liouvillian (see Appendix \ref{A:BosonTwoLevelCoefficients}), following the method outlined in \secref{S:ThoereticalFramework}. In the wide-band limit with $\Gamma_\nu(\omega_n)=\Gamma_\nu$, we find that these currents explicitly read as
\begin{align}
  \dot{N}_\nu &=-\sum_{m=1}^2 m \Gamma_{\nu}[\bar n_\nu^{(+)}(\omega_{m})(\rho_{m}-\rho_{m-1})+\rho_{m}],\label{E:BoseTwoLevelCurrent}\\
  \dot{E}_\nu &=-\sum_{m=1}^2 m \Gamma_{\nu}[\bar n_\nu^{(+)}(\omega_{m})(\rho_{m}-\rho_{m-1})+\rho_{m}] \omega_m\label{E:BoseTwoLevelEnergyCurrent}.
\end{align}

In order to compare the bosonic evolution to the fermionic one, we consider reservoir temperatures above $T_C$ such that we can neglect the ground-state contribution in \equref{E:AverageParticleNumber}. Consequently, the equation of motion for the thermodynamic variables of the reservoirs is given by \equref{E:ExplicitSystem}, and we can numerically calculate their evolution. The results for the bosonic transport system with two, and also with one transport channel are in qualitative agreement with our fermionic results.

As an example, we present in \figref{fig::BoseSingleDirCurrent}(a) the dynamics of the temperature difference between the bosonic reservoirs for a quantum system with a single transition energy. Analogous to the fermionic results shown in \figref{fig::FerSingleDirCurrent}, we find that the steady state in this setup is nonthermal. Here, the initial particle-number bias is converted into a steady-state temperature bias, whose amount and sign can be tuned by shifting the transition energy $\omega$ of the quantum system, as can be seen in \figref{fig::FerSingleDirCurrent}(b). For this initial non-equilibrium configuration, we find from \equref{E:FullEoMSolutionN} that the steady-state temperature bias is given by
\equ{
    \Delta_T(\infty) \equiv \lim_{t\rightarrow \infty}\Delta_T(t)= -\frac{\sigma (\mu_{\rm{eff}}-T\Sigma) }{\Omega N^2 \kappa C} \Delta_N{(0)}.
    }

The above linear transport coefficients have to be derived from the linearized steady-state currents running through the quantum system \cite{Nietner2013}.
%
%
\section{Summary}\label{S:Summary}
%
%
In this paper, we analyzed the equilibration process between two reservoirs, which are initialized in a nonequilibrium configuration and that are weakly thermally connected via a few-level quantum system. To this end, we established the full equations of motion describing the evolution of the density matrix elements of the quantum system, as well as the evolution of the thermodynamic variables of the attached reservoirs. Subsequently, these equations were solved, both numerically and analytically, by a linearized theory. We observe a qualitative dependence of the equilibration on the number of available transport channels. Only setups with more than one accessible transport channel show a thermodynamic equilibration for long times, whereas a nonthermal steady state is reached in systems with only a single transport channel. This fundamentally different behavior might be used to construct a transistor or capacitor for ultracold atoms. Such a machine would also work quite efficiently, as we confirmed from the 
calculation of the heat current and power output. Finally, we compare the equilibration process in thermal fermionic and bosonic transport setups, where we qualitatively observe the same behavior.

\section{Acknowledgements}
We gratefully acknowledge the financial support by the DFG (Grants No.~SFB 910, No.~BR 1528/7-1,8-2,9-1 and No.~GRK 1558). Furthermore, we thank S.~Krinner for inspiring and helpful discussions.
\appendix
\begin{widetext}
\section{Atoms confined in a 3D cubic box}\label{A:Box}
%
%
In the case of an ideal noninteracting quantum gas confined in a 3D cubic box of volume $V$ with periodic boundary conditions, the dispersion relation reads $\varepsilon_k=k^2/(2m)$. In the thermodynamic limit, one can obtain the total particle number $N_\nu=\sum_k\bar{n}_\nu^{(\xi)}(\varepsilon_{k})$ and energy $U_\nu=\sum_k\varepsilon_k\bar{n}_\nu^{(\xi)}(\varepsilon_k)$  by substituting the summation by an integral
\begin{equation}
  \frac{1}{(2\pi)^3}\sum_k\rightarrow \int_0^\infty g(\varepsilon)d\varepsilon ,\ \quad g(\varepsilon)=\frac{2\pi Vg_s}{(2\pi)^3}(2m)^{\frac{3}{2}}\varepsilon^{\frac{1}{2}},
\end{equation} 
where $g_s=(2S+1)$ is the spin degeneracy coefficient. Integrating these expressions, one finds
\begin{equation}
  N_\nu=g_s V\frac{\xi}{\lambda}\text{Li}_{\frac{3}{2}}(\xi z_\nu)+g_s N_\nu^{(0)}(\xi),\ \quad U_\nu=\frac{3}{2}g_s V\frac{\xi}{\lambda_\nu}T_\nu\text{Li}_{\frac{5}{2}}(\xi z_\nu)
\end{equation}
with $\lambda=\sqrt{2\pi/(m T_\nu)}$ being the thermal wavelength of particles with mass $m$. Subsequently, one can derive the characteristic temperature scales for bosons and fermions which read as
\begin{equation}
  T_C=\frac{2\pi}{m}\left(\frac{N_\nu}{g_s V\zeta(3/2)}\right),\ \quad T_F=\frac{1}{2m}\left(\frac{6\pi^2N_\nu}{g_sV}\right),
\end{equation} 
where $\zeta(x)$ is the Riemann zeta.
From this consideration, we see that assuming different boundary conditions for the reservoirs leads to modified energy scales and different poly-logarithms. However, from numerical calculations we find that the qualitative dynamical properties are the same as those discussed in \secref{S:Results}.
%
%
\section{Fermionic Liouvillian}\label{A:Liouvillian}
%
%
The Liouvillian of a noninteracting fermionic system with $l$ channels and energy independent tunneling rates $\Gamma_\nu(\omega)=\Gamma_\nu$, reads as
{
\begin{align}
  \mathcal{L}^{(\nu)}= \frac{\Gamma_\nu}{2} \left[\begin{array}{ccccc}
  -\bar n^{(-)}_\nu(\omega_1)-\ldots -\bar n^{(-)}_\nu(\omega_l)&1-\bar n^{(-)}_\nu(\omega_1)&1-\bar n^{(-)}_\nu(\omega_2)&\ldots&1-\bar n^{(-)}_\nu(\omega_l)\\
  \bar n^{(-)}_\nu(\omega_1)&-1+\bar n^{(-)}_\nu(\omega_1)&0&\ldots&0\\
  \bar n^{(-)}_\nu(\omega_2)&0&-1+\bar n^{(-)}_\nu(\omega_2)&\vdots&\vdots\\
  \vdots&\vdots&\ldots&\ddots&0\\
  \bar n^{(-)}_\nu(\omega_l)&0&\ldots&0&-1+\bar n^{(-)}_\nu(\omega_l)\\\end{array}\right]\label{eq:A:LiuF},
\end{align}}
with the corresponding reduced system density matrix given by $\rho=(\rho_0,\rho_1,\ldots,\rho_l)^{\rm{T}}$, where $\rho_0$ is the ground-state population, and $\rho_j$ is the population of the $j$th excited state. 
From this Liouvillian, we can calculate the evolution of the reduced system density matrix from \equref{eq::rhofermi}. Moreover, using \equref{eq:A:LiuF} together with the Eqs.~\eqref{E:StaticCurrentDefinition} and \eqref{E:StaticEnergyCurrentDefinition} allows to calculate the energy and particle currents running through the quantum system. 

Considering a system with a single transition energy $\omega_1$, the above Liouvillian is truncated at $\omega_1$ resulting in a $2\times2$ matrix, and we find for the rate matrix
{
\begin{align}
  \mathcal{L}^{(\nu)} &= \frac{\Gamma_\nu}{2} \left[
  \begin{array}{ccccc}
    0 & [1-\bar n^{(-)}_\nu(\omega_1)]\\
    -\bar n^{(-)}_\nu(\omega_1) & 0\\
  \end{array}\right] ,\label{E:AppendixOneLevelDevChi}
\end{align}}
which results with  $\rho=(1-\rho_1,\rho_1)^{\rm{T}}$ in the particle and energy currents given in Eqs.~\eqref{E:FermiSingelLevelCurrents}.
Analogously, we find for fermionic systems with two transition energies that the rate matrix reads as
{
\begin{align}
  \mathcal{L}^{(\nu)} &= \frac{\Gamma_\nu}{2} \left[
  \begin{array}{ccccc}
    0 & [1-\bar n^{(-)}_\nu(\omega_1)] & [1-\bar n^{(-)}_\nu(\omega_2)]\\
    -\bar n^{(-)}_\nu(\omega_1) & 0 & 0\\
    -\bar n^{(-)}_\nu(\omega_2) & 0 & 0
  \end{array}\right],
\end{align}}
which results with  $\rho=(\rho_0,\rho_1,\rho_2)^{\rm{T}}/{\rm{Tr}\{\rho\}}$ in the particle and energy currents given in Eqs.~\eqref{E:FermiTwoLevelCurrent} and \eqref{E:FermiTwoLevelEnergyCurrent}.
%
%
\section{Single Fermionic Transport-Channels}\label{A:FermiOneLevelCoefficients}
%
%
For an irreducible rate matrix, the equation $0 = \sum_{i,\nu}\mathcal L^{(\nu)}_{i,j}\bar\rho_j $ together with the normalization $\tr{\bar\rho}=1$, uniquely determines the steady-state reduced system density matrix $\bar\rho$. 
For a system with a single transition energy $\omega$, the Liouvillian in \equref{eq:A:LiuF} is truncated at $\omega=\omega_1$. The corresponding steady-state density matrix is given by $\bar\rho=(1-\bar \rho_1 , \bar \rho_1)^{\rm{T}}$ with $\bar\rho_1= \rklamm{\Gamma_L \bar n_L^{(-)}+\Gamma_R \bar n_R^{(-)}}/(\Gamma_L+\Gamma_R)$.

Inserting Eq.~\eqref{E:AppendixOneLevelDevChi}  and the steady-state density matrix into the respective current equations \eqref{E:StaticCurrentDefinition} and \eqref{E:StaticEnergyCurrentDefinition}, yields the steady-state currents $J_N=J_N^{(L)}=-J_N^{(R)}$ and $J_E=J_E^{(L)}=-J_E^{(R)}$, which read as
\begin{align}
  J_N = \frac{\Gamma_L \Gamma_R}{\Gamma_L + \Gamma_R} \eklamm{  \bar n^{(-)}_L(\omega ) -\bar n^{(-)}_R(\omega )}\hspace{.5em}{\rm{and}}\hspace{.5em}
  J_E = \omega J_N.
\end{align}

Subsequently, constructing the linear-response heat current $J_Q= J_E-(\mu+T\alpha)J_N$ \cite{Nietner2013}, and linearizing the heat and particle current with respect to the affinities $\Delta_T$ and $\Delta_N/(N^2 \kappa)$, results in the linear-response transport coefficients 
\equ{
    \sigma = \frac{\Gamma_L \Gamma_R}{T(\Gamma_L+\Gamma_R)} [1-\bar n^{(-)}(\omega)] \bar n^{(-)}(\omega),\hspace{1em}q=0,\hspace{1em}\Sigma = \frac{1}{T}\left(\mu +\alpha  T-\omega \right).
    }

In the single-transport-channel situation, energy and particle currents are proportional and consequently the heat conductance $q$ vanishes. The result for the particle conductance $\sigma$ is consistent with the well-known Coulomb blockade conductance peak for a single resonant level \cite{Beenakker1991}.
%
%
\section{Two Fermionic Transport-Channels}\label{A:FermiTwoLevelCoefficients}
%
%
In case of a system with two transport channels, the Liouvillian from \equref{eq:A:LiuF} is truncated at $\omega_2$ resulting in a $3\times3$ matrix. Solving the respective current equations from \equref{E:StaticCurrentDefinition} and \equref{E:StaticEnergyCurrentDefinition} and inserting the corresponding steady-state density matrix, yields the steady-state currents $J_N=J_N^{(L)}=-J_N^{(R)}$ and $J_E=J_E^{(L)}=-J_E^{(R)}$. For simplicity, we suppose that the rates are energy independent and homogeneous, \ie, $\Gamma_\nu(\omega)=\Gamma_\nu=\Gamma$, which results in the expressions
\begin{gather}
  J_N = \Gamma\frac{  \bar n^{(-)}_L(\omega _1) \left[1-\bar n^{(-)}_L(\omega _2)\right]-\bar n^{(-)}_R(\omega _1) \left[1-\bar n^{(-)}_R(\omega _2)\right]+\bar n^{(-)}_L(\omega _2)-\bar n^{(-)}_R(\omega _2)}{ \left[\bar n^{(-)}_L(\omega _1)+\bar n^{(-)}_R(\omega _1)\right] \left[\bar n^{(-)}_L(\omega _2)+\bar n^{(-)}_R(\omega _2)\right]-4},\\
  J_E = \frac{\Gamma }{2}\frac{\sum_{j=1}^2 \omega_j \left[2-\bar n^{(-)}_L(\omega _j)-\bar n^{(-)}_R(\omega _j)\right] \left[\bar n^{(-)}_L(\omega _j)-\bar n^{(-)}_R(\omega _j)\right]}{\left[\bar n^{(-)}_L(\omega _1)+\bar n^{(-)}_R(\omega _1)\right] \left[\bar n^{(-)}_L(\omega _2)+\bar n^{(-)}_R(\omega _2)\right]-4}.
\end{gather}

Subsequently, we construct the linear-response heat current $J_Q= J_E-(\mu+T\alpha)J_N$ \cite{Nietner2013}, and linearize the heat and particle currents, with respect to their affinities $\Delta_T$ and $\Delta_N/(N^2 \kappa)$. Applying the definitions from \equref{E:sigmaDefinition}, \equref{E:heatDefinition} and \equref{E:SeebeckDefinition}, results in the respective linear-response transport coefficients 
\begin{gather}
    \sigma = \frac{\Gamma}{4 T}\frac{[1-\bar n^{(-)}(\omega_1)][1-\bar n^{(-)}(\omega_2)][\bar n^{(-)}(\omega_1)+\bar n^{(-)}(\omega_2)]}{1-\bar n^{(-)}(\omega_1)\bar n^{(-)}(\omega_2)},\hspace{.8em}
    q = \sigma\frac{\bar n^{(-)}(\omega_1)\bar n^{(-)}(\omega_2)(\omega_1-\omega_2)^2}{T[\bar n^{(-)}(\omega_1)+\bar n^{(-)}(\omega_2)]^2},\\
    \Sigma = \frac{\sum_{j=1}^2 \bar n^{(-)} (\omega _j) \left(\mu +\alpha  T-\omega _j\right)}{T\left[\bar n^{(-)} (\omega _1)+\bar n^{(-)} (\omega _2)\right]}.
\end{gather}

Here, we find a finite heat conductance $q\neq0$, which is proportional to the difference $\omega_1-\omega_2$ of the transition energies. It allows for a full equilibration of the reservoirs.
%
%
\section{Two Bosonic Transport-Channels}\label{A:BosonTwoLevelCoefficients}
%
%
Starting from the system Hamiltonian in \equref{E:BoseHamiltonian} and following the procedure outlined in \secref{S:MasterEquation}, we obtain the following Liouvillian with energy-independent rates $\Gamma_\nu(\omega)=\Gamma_\nu$ for the bosonic system with at most two particles
\begin{align}
  \mathcal{L}^{(\nu)}= \frac{\Gamma_\nu}{2} \left[
  \begin{array}{ccccc}
    -\bar n^{(+)}_\nu(\omega_1) & 1+\bar n^{(+)}_\nu(\omega_1) & 0\\
    \bar n^{(+)}_\nu(\omega_1) & -1-\bar n^{(+)}_\nu(\omega_1)-2\bar n^{(+)}_\nu(\omega_2) &  2+2\bar n^{(+)}_\nu(\omega_2)]\\
    0 &  2\bar n^{(+)}_\nu(\omega_2) & -2-2\bar n^{(+)}_\nu(\omega_2)]\\
  \end{array}\right].\label{eq:A:LiuB}
\end{align}

Subsequently, from the definitions in \equref{E:StaticCurrentDefinition} and in \equref{E:StaticEnergyCurrentDefinition}, we derive the particle and energy currents presented in \equref{E:BoseTwoLevelCurrent} and \equref{E:BoseTwoLevelEnergyCurrent}.
\end{widetext}
%
%
 \bibliographystyle{apsrev}
%

\end{document}